\documentclass[preprint,aps]{revtex4}
\usepackage{graphicx}
\usepackage{bm}
\begin{document}

\preprint{UA-NPPS/03/2005}
\title{Critical QCD in Nuclear Collisions}

\author{N.G. Antoniou} 
\affiliation{Nuclear Theory Group, Brookhaven National Laboratory,
Upton, NY 11973, USA}
\altaffiliation[On leave of absence from: ]{Department of Physics, University of Athens, GR-15771 Athens, 
Greece}
\author{Y.F. Contoyiannis}
\author{F.K. Diakonos}
\email{fdiakono@cc.uoa.gr}
\author{G. Mavromanolakis}
\affiliation{Department of Physics, University of Athens, GR-15771 Athens, 
Greece}

\date{\today}

\begin{abstract}
A detailed study of correlated scalars, produced
in collisions of nuclei and associated with the $\sigma$-field
fluctuations, $(\delta \sigma)^2= < \sigma^2 >$, at the QCD critical
point (critical fluctuations), is performed on the basis of a
critical event generator (Critical Monte-Carlo) developed in 
our previous work. The aim of this analysis is to reveal 
suitable observables
of critical QCD in the multiparticle environment of simulated events
and select appropriate signatures of the critical point, associated
with new and strong effects in nuclear collisions.
\end{abstract}

\pacs{25.75.Nq,12.38.Mh,25.75.-q}

\maketitle

\section{Introduction}
The existence of a critical point in the phase diagram of QCD, for
nonzero baryonic density, is of fundamental significance for our
understanding of strong interactions and so its experimental verification
has become an issue of high priority \cite{WRB}. For this purpose an 
extensive programme of event-by-event searches for critical fluctuations 
in the pion sector is in progress in experiments with heavy ions from SPS to
RHIC energies \cite{NA49}. In \cite{ACDKK} we have emphasized, however, 
that in order to reveal critical density fluctuations in multiparticle
environment, one has to look for unconventional properties in the momentum
distribution of reconstructed dipions ($\pi^+ \pi^-$-pairs) 
\footnote{We use the term dipions
to indicate exclusively a pair of opposite charged pions. This will be the case throughout 
in the present work.} 
with invariant
mass just above the two-pion threshold. In fact, the QCD critical point, if it exists, 
communicates with a zero mass scalar field ($\sigma$-field) which at lower
temperatures ($T < T_c$) may reach the two-pion threshold and decay
in very short time scales owing to the fact that its coupling to the two-pion system 
is strong. Obviously, the fundamental, underlying pattern of $\sigma$-field 
fluctuations, built-up near the critical point by the universal critical exponents
of QCD \cite{ACDKK}, is phenomenologically within reach if and only if
the study of correlated sigmas, reconstructed near the two-pion
threshold, becomes feasible. In the present work we perform a detailed
feasibility study of the observables related to the detection
of the QCD critical point in nuclear collisions. In order to proceed 
we summarize, first, the principles on which the behaviour
of a critical system of sigmas is based \cite{ACDKK,WRB}. 

(a) The geometrical structure  of the critical system in transverse
space (after integrating in rapidity) consists of $\sigma$-clusters 
with a fractal dimension $d_F=\frac{2(\delta-1)}{\delta+1}$
leading to a power law, $< \sigma^2> \sim \vert \vec{x} \vert^{-
\frac{4}{\delta+1}}$, for the $\sigma$-field fluctuations, within each
cluster ($\delta$ : isotherm critical exponent)

(b) In transverse momentum space the $\sigma$-fluctuations obey a
power law $< \sigma^2> \sim \vert \vec{p}_{\perp} \vert^{-\frac
{2(\delta-1)}{\delta+1}}$ leading to observable intermittent
behaviour of factorial moments: $F_2(M) \sim (M^2)^{\frac{\delta-1}
{\delta+1}}$ where $M^2$ is the number of $2D$ cells \cite{BialPes}.

(c) The density fluctuations of dipions ($\pi^{+}\pi^{-}$-pairs) with
invariant mass close to two-pion threshold $(2 m_{\pi})$ incorporate
the sigma-field fluctuations, $(\delta \sigma)^2
\approx <\sigma^2>$, at the critical point, under the assumption that
the sigma mass reaches the two-pion threshold $(m_{\sigma} 
\stackrel{>}{\sim} 2 m_{\pi})$ in a time scale shorter than the relaxation
time of critical fluctuations.

(d) The QCD critical point belongs to the universality class of the
$3D$ Ising system in which $\delta \approx 5$.

On the basis of these principles and the fact that critical
clusters in the above universality class interact weakly \cite{ACDPRE}, 
one may construct a Monte-Carlo generator (Critical Monte-Carlo: CMC) able to
simulate events of critical sigmas, correlated according to the above prescription 
\cite{ACDKK}. Then, if the corresponding invariant mass distribution is known, we can also 
simulate the decay of the critical sigmas to pions which are experimentally observable.
To complete the numerical experiment we use the momenta of the charged pions, obtained
from the decay of the critical sigmas, to form, in an event-by-event basis, neutral dipions 
($\pi^+ \pi^-$ pairs) and to look for fingerprints of the original critical sigma fluctuations
in their momentum distribution. To check the applicability of our approach we perform also a 
comparative study between the revealed critical correlations-fluctuations in CMC and the 
corresponding behaviour of a conventional Monte-Carlo (HIJING).

The input parameters of the simulation are the size of the system in rapidity ($\Delta$) and 
transverse space ($R_{\perp}$) as well as the proper time scale ($\tau_c$) characteristic for 
the formation of the critical system. In what follows we will use exclusively $\Delta=6$ for the 
rapidity size, adapted to the SPS-NA49 conditions ($E_{beam} \approx 158~GeV/n$). The parameters $R_{\perp}$, 
$\tau_c$ can then be tuned in order to fit the mean charged pion multiplicity of systems of different
size ($CC$, $SiSi$ and $PbPb$) studied experimentally at this energy \cite{NA49AA}. For each choice of 
input parameters we produce, using the CMC generator,
a large set of critical sigma events. In order to generate the corresponding pion sector we assign to the
sigmas the invariant mass distribution $\rho(m_{\sigma})$ and then we let them decay into pions with 
a branching ratio $1:2$ for neutral to charged. The choice of $\rho(m_{\sigma})$ is determined by the 
requirement that the resulting inclusive neutral dipion invariant mass distribution $\rho(m_{\pi^+ \pi^-})$ 
resembles to a large extent the corresponding distribution obtained in experiments with heavy ion collisions
at high energies. In order to perform our analysis in terms of observable quantities we choose the 
mean multiplicity of positive charged pions per event $<n_{\pi^+}>$ as the
basic parameter (instead of $R_{\perp}$, $\tau_c$) characterizing different $A+A$ systems. We note that 
within the CMC approach the property $<n_{\pi^+}>=<n_{\pi^-}>$ is exactly fullfiled. We will investigate 
three different cases: low ($<n_{\pi^+}>\approx 10$), intermediate ($<n_{\pi^+}> \approx 30$) and high 
($<n_{\pi^+}> \approx 220$) positive charged pion multiplicity resembling the $C+C$, $Si+Si$ and $Pb+Pb$ system at 
$158~GeV/n$ respectively.

\section{The critical sigma sector}
First we consider the case of low pion multiplicity. Before going on with the analysis it is worth to emphasize
that, in general, the determination of the critical sigma sector using the observed momenta of charged pions, 
is a very difficult task. This is due to the absence of a characteristic pattern in the inclusive dipion 
invariant mass distribution attributed to the presence of critical sigmas. To illustrate this property we show
in Fig.~1 the inclusive distribution of the sigma invariant mass calculated before the decay of the 
sigmas into pions (solid line) as well as the corresponding distribution for neutral dipions recontsructed from 
the final pion momenta (full circles) for $30000$ CMC events with $<n_{\pi^+}>=11.31$ (dataset I). Both distributions
are equally normalized. Apart from a peak at $m_{\pi^+ \pi^-} \approx 450~MeV$ due to kinematics there is no other 
structure in the invariant mass profile. Therefore the detection of the critical sigma sector has to go through the, 
more subtle, study of density fluctuations in momentum space. 

\begin{figure}
\includegraphics{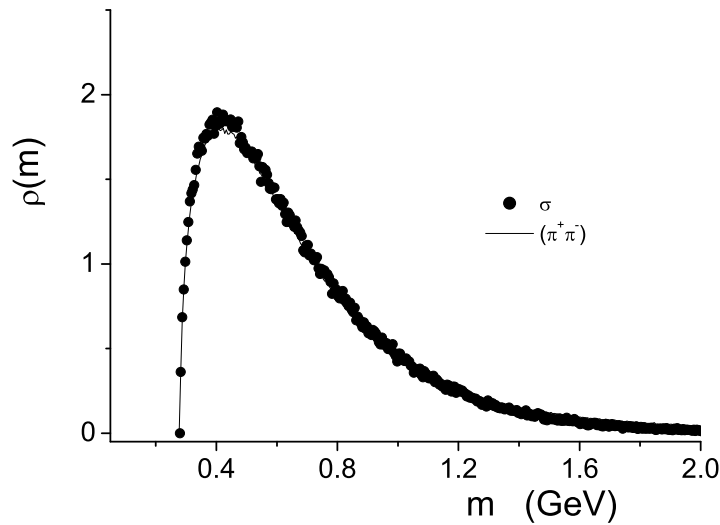}
\caption{\label{fig:fig1} The critical sigma invariant mass distribution $\rho(m_{\sigma})$ (solid line)
as well as the corresponding neutral dipion distribution $\rho(m_{\pi^+ \pi^-})$ (full circles) for
$30000$ CMC events with $<n_{\pi^+}>=11.31$.}
\end{figure}

In order to reveal the underlying critical fluctuations, at the level of observation, 
one has first to perform factorial moment analysis in small cells of the momentum space \cite{BialPes}. We have 
chosen transverse momenta for this analysis in order to avoid additional assumptions about the role
of longitudinal rapidity in the description of the statistical mechanics of the system \cite{ACDKK}. Applying 
factorial moment analysis 
to the transverse momenta $(p_x,p_y)$ of the negative pions in the sample of the $30000$ critical events we obtain 
for the second moment a weak intermittency effect: $F_2 \sim M^{2 s_2}$ with $s_2 \approx 0.077$ much smaller than 
the expected to occur in the critical system ($s_2 = 2/3$). We note with $M$ the number of bins in each 
momentum space component. The corresponding factorial moment for the sigmas, before their decay, follows closely 
the theoretical prediction: $s_2 \approx 0.66$. This behaviour is displayed in Fig.~2. 

\begin{figure}
\includegraphics{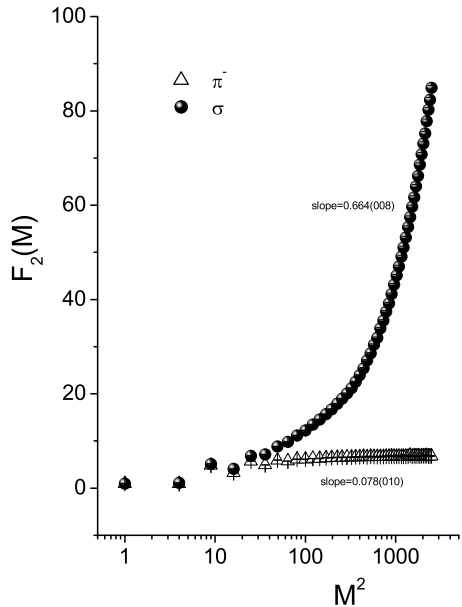}
\caption{\label{fig:fig2} The second factorial moment $F_2$ in transverse momentum space
of both the negative pions as well as the critical sigmas. We use the same dataset as in Fig.~1.}
\end{figure}

The reason for the suppression of fluctuations in the pionic sector is the kinematical distortion of the 
self-similar pattern formed in the sigma sector due to the sigma-decay. The strength of this distortion
increases with momentum transfer: $Q=\sqrt{m^2_{\sigma}-4m_{\pi}^2}$ and becomes negligible near the two-pion 
threshold ($m_{\sigma} \approx 2 m_{\pi}$). Here $m_{\sigma}^2=(p_{\pi^+}+p_{\pi^-})^2$ where $p_{\pi^{\pm}}$ 
are the four momenta of the charged pions produced through the sigma-decay. Thus the search for critical
fluctuations is based on an accurate reconstruction of the momenta of the decaying sigmas. In particular
$(\pi^+,\pi^-)$ pairs with invariant mass close to the two-pion threshold are the best candidates to carry
potentially the geometrical features of the critical isoscalars. The great advantage of our approach is
that it allows a self-consistency test for the reconstruction of the sigma sector at the level of density
fluctuations due to our exact knowledge of the sigma momenta and the corresponding fluctuation pattern.

\section{Reconstruction of power-laws}
In practice the reconstruction of the critical momentum fluctuations is performed by looking, event by event, 
for $(\pi^+,\pi^-)$ pairs fullfiling the criterion A:
\begin{equation}
A=\{(\pi^+,\pi^-) \vert 4 m_{\pi}^2 \leq (p_{\pi^+} + p_{\pi^-})^2 \leq 
(2 m_{\pi} + \epsilon)^2 \}
\label{eq:eq1}
\end{equation} 
with $\epsilon \ll 2 m_{\pi}$.
The momentum of the corresponding neutral dipion is then obtained as:
$\vec{p}_{\pi\pi}=\vec{p}_{\pi^+} + \vec{p}_{\pi^-}$.
In order to ensure that {\bf all} the available critical sigmas within the above kinematical region are 
recovered in the reconstruction we have to use {\bf full} pairing forming {\bf all} possible pairs 
$(\pi^+,\pi^-)$ fullfiling (\ref{eq:eq1}) for a given $\pi^+$. However the full pairing introduces
as a side effect a combinatorial background which has to be treated appropriately. We will come back to this 
point later on. Thus for each value of $\epsilon$ we obtain a set of events including dipion momenta. 
We perform factorial moment analysis in transverse momentum space for each such dataset. According to the
previous discussion, for decreasing values of $\epsilon$ the fluctuations measured by the intermittency 
exponent $s_2$ of the corresponding factorial moment should increase leading to $s_2 \to \frac{2}{3}$
as $\epsilon \to 0^+$. In Fig.~3a we show the second moment in transverse momentum space obtained from 
data sets of reconstructed dipions for three different values of $\epsilon$ ($5,50$ and $500~MeV$).   
We observe an increase of the slope $s_2$ with decreasing $\epsilon$. However this increase has two different
origins: (i) the presence of nontrivial fluctuations of dynamical origin and (ii) kinematically induced
fluctuations through the constraint (\ref{eq:eq1}) which become of the same order as the dynamical
ones for $\epsilon \to 0$. It must be noted that the functional forms of $F_2$ for the different $\epsilon$
values are not exact power-laws and the exponent $s_2$ is an effective one. Increasing $\epsilon$ the deviations 
from a power-law description increase too. In order to control the kinematical fluctuations we form datasets 
with mixed events
through appropriate shuffling of the momenta in the original CMC events. Then we calculate the second factorial
moments for the datasets consisting of mixed events. We show in Fig.~3b the results of this analysis. Also in the
case of mixed events we observe a clear increase of the effective slope $s_2^{(m)}$ \footnote{The superscript $(m)$ 
is used to indicate mixed events} for decreasing $\epsilon$. Due to the absence of dynamical fluctuations in the mixed 
events it is natural to conclude that the observed behaviour in $F_2^{(m)}$ is attributed to the kinematical 
fluctuations.

\begin{figure}
\includegraphics{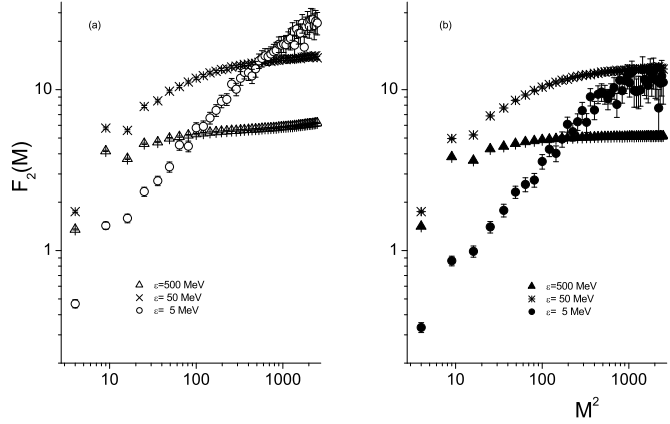}
\caption{\label{fig:fig3} (a) The second factorial moment $F_2$ in transverse momentum space
of reconstructed dipions for $\epsilon=5~,50$ and $500~MeV$ (log-log plot). We use the same 30000 CMC 
events as in Fig.~1. In (b) we show for comparison the corresponding factorial moments for datasets
consisting of mixed events.} 
\end{figure}

These can be removed if we introduce the difference $\Delta F_q=F_q-F_q^{(m)}$ between the 
factorial moments of real and mixed events (for any order $q$). For $q=2$ the quantity $\Delta F_2$ represents
the correlator of the corresponding dipions. Thus in the difference $F_2-F_2^{(m)}$ the combinatorial background
representing the uncorrelated part of the reconstructed dipions is suppressed and the critical fluctuations
associated with the correlated part of the dipion sector are recovered. We illustrate how this subtraction 
succeeds in practice in Fig.~4 where we present $\Delta F_2$ for $\epsilon=5~MeV$ and $<n_{\pi^+}>=11.31$. 
Although the moments $F_2$ and $F_2^{(m)}$ are not exact power-laws, as indicated above, the difference 
$\Delta F_2$ is significantly better fitted by a power-law and the corresponding slope $\phi_2=0.65(02)$ is very 
close to the slope obtained from the dataset of the sigmas before their decay (see Fig.~1).   

\begin{figure}
\includegraphics{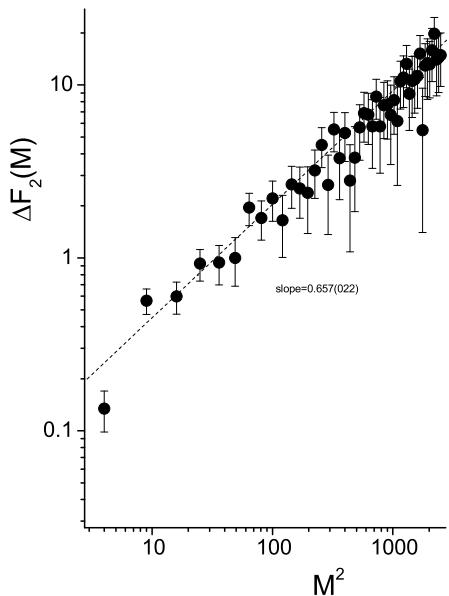}
\caption{\label{fig:fig4} (a) The log-log plot of the correlator $\Delta F_2$ in transverse momentum space
of reconstructed dipions for $\epsilon=5~MeV$ obtained using 30000 CMC events (the same as in Fig.~1)
and the corresponding mixed events.} 
\end{figure}

The fluctuations in the CMC events, as described by $\Delta F_2$, are induced by strongly correlated critical QCD dynamics
and are expected to be at least one order of magnitude greater than fluctuations originating from conventional hadronic dynamics. 
To check the validity of this statement we have calculated $\Delta F_2$ using $33176$ events obtained from the HIJING Monte-Carlo 
generator \cite{Hijing} simulating the SPS $C+C$ system at $158~GeV/n$ and involving only noncritical QCD dynamics. 
In Figs.~5a-d we show the results of our calculations for four different values of $\epsilon$. For comparison we plot in the same 
figure the correlator for the $30000$ CMC events with $<n_{\pi^+}>=11.31$. We observe that for all values of $\epsilon$ 
the correlator $\Delta F_2$ for the HIJING system remains flat fluctuating around zero, while for the CMC events as $\epsilon$ 
decreases the characteristic, for the critical system, power-law behaviour is setting on according to the previous discussions.
  
\begin{figure}
\includegraphics{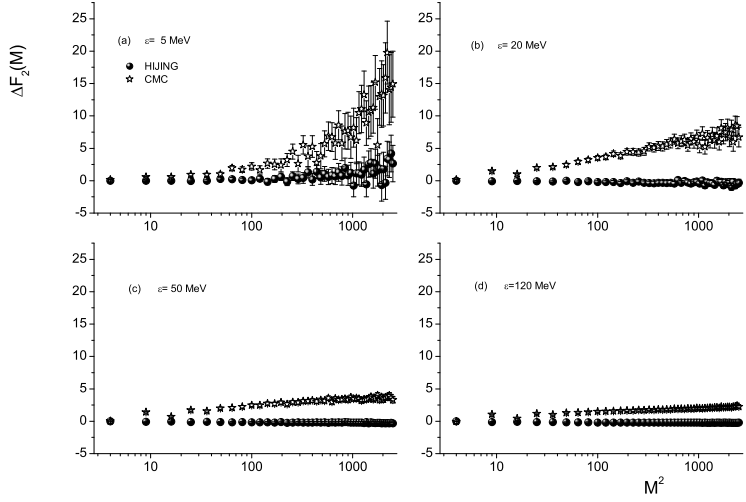}
\caption{\label{fig:fig5} The correlator $\Delta F_2$ for the CMC and the HIJING system using four different values of $\epsilon$:
(a) 5~MeV, (b) 20~MeV, (c) 50~MeV and (d) 120~MeV.} 
\end{figure}

\section{The critical index}
It is worth investigating in more detail the role of the parameter $\epsilon$ in our analysis. Practically the
desired limit $\epsilon \to 0+$ is not accessible for several reasons. First of all looking at Fig.~1 we observe
that in this limit the number of initial critical sigmas as well as the number of reconstructed dipions vanishes.
We practically need infinite statistics in order to extend our analysis in this kinematical region. In addition,
in the range of very small $\epsilon$ values the fluctuations induced by the kinematical constraint (\ref{eq:eq1})
become of the same order as the dynamical fluctuations due to the critical sigmas. Our entire treatment is based on
an accurate cancellation of the kinematical fluctuations in the difference $\Delta F_2$ and this requires again very high 
statistics if $\epsilon \to 0$. In fact, using $\epsilon$ values smaller than $4~MeV$ in the analysis of the $30000$ CMC events
considered so far, the statistical errors increase rapidly and prohibit a reliable reveal of the critical fluctuations. 
On the other hand, if we increase $\epsilon$, the combinatorial background due to the 
reconstructed dipions is increasing rapidly and the relative measure of 
the critical sector goes to zero. Therefore we must search for an optimal region of $\epsilon$ values to
perform our analysis. In order to achieve a more quantitative criterion for the determination of the appropriate region
of $\epsilon$-values we consider more carefully the sector of reconstructed dipions. In the CMC case the reconstructed dipions 
are divided into two
subsets: the set of real sigmas and the set of fake sigmas. Let us denote by $<n_{r,\sigma}>_{\epsilon}$ the mean number of real 
sigmas per event with invariant mass in the kinematical window (\ref{eq:eq1}) for a given value of $\epsilon$. Within our approach 
we investigate only the part of the sigmas which decays into a pair of opposite charged pions. Therefore we have:
\begin{equation} 
<n_{r,\pi^+}>_{\epsilon}=<n_{r,\pi^-}>_{\epsilon}=<n_{r,\sigma}>_{\epsilon}
\label{eq:eq2}
\end{equation}
where $<n_{r,\pi^{+(-)}}>_{\epsilon}$ is the mean number of positive (negative) charged pions per event produced through the decay 
of the real sigmas with invariant mass in the region (\ref{eq:eq1}). Obviously to a good approximation the corresponding   
number of fake sigmas is given by:
\begin{equation}
<n_{f,\sigma}>_{\epsilon} \approx <n_{r,\pi^+} n_{r,\pi^-}>_{\epsilon} - <n_{r,\pi^+}>_{\epsilon} 
\label{eq:eq3}
\end{equation}
The first term on the right hand side of eq.(\ref{eq:eq3}) is dominated by uncorrelated $(\pi^+,\pi^-)$ pairs. Therefore we can
write: $<n_{r,\pi^+} n_{r,\pi^-}>_{\epsilon} \approx <n_{r,\pi^+}>_{\epsilon} <n_{r,\pi^-}>_{\epsilon}$ und using the property:
$<n_{r,\pi^+}>_{\epsilon}=<n_{r,\pi^-}>_{\epsilon}$ we finally obtain:
\begin{equation}
<n_{f,\sigma}>_{\epsilon} \approx <n_{r,\pi^+}>^2_{\epsilon} - <n_{r,\pi^+}>_{\epsilon} 
\label{eq:eq4}
\end{equation}
A natural constraint ensuring the dominance of the real sigmas over the fake ones is to use in our analysis $\epsilon$ values 
fullfiling the condition:
\begin{equation}
<n_{f,\sigma}>_{\epsilon}~<~<n_{r,\sigma}>_{\epsilon} ~~ \Rightarrow ~~ <n_{r,\pi^+}>^2_{\epsilon} - <n_{r,\pi^+}>_{\epsilon}~<~
<n_{r,\pi^+}>_{\epsilon} 
\label{eq:eq5}
\end{equation}
which simplifies to: $<n_{r,\pi^+}>_{\epsilon}~<~2$. This is fullfiled for example for the value $\epsilon=5~MeV$ which we have used
so far in our analysis since in this case $<n_{\pi^+ \pi^-}>(5~MeV)=1.12$. We also observe that $<n_{\pi^+ \pi^-}>_{\epsilon}~>~
<n_{r,\pi^+}>_{\epsilon}$ for any value of $\epsilon$, where $<n_{\pi^+ \pi^-}>_{\epsilon}$ is the mean number of dipions per event 
obtained through the reconstruction 
in the kinematical window (\ref{eq:eq1}). The upper bound in $<n_{r,\pi^+}>_{\epsilon}$ corresponds to an upper bound for $\epsilon$. 
The lower bound both 
for $\epsilon$ as well as $<n_{r,\pi^+}>_{\epsilon}$ is determined by the statistics according to the discussion in the previous paragraph. 
For practical purposes one applies the stronger bound $<n_{\pi^+ \pi^-}>_{\epsilon}~<~2$ for the estimation of the appropriate region of 
$\epsilon$-values to be used in the data analysis. The great advantage of restricting $<n_{\pi^+ \pi^-}>_{\epsilon}$ instead of $\epsilon$ becomes 
more clear when we compare the analysis in CMC datasets simulating $A+A$ processes of different size. In this case the dependence of 
$<n_{\pi^+ \pi^-}>_{\epsilon}$ on $\epsilon$ varies from system to system. To illustrate this we have produced two additional CMC datasets, 
with $<n_{\pi^+}>=29.69$ (dataset II) and $<n_{\pi^+}>=213.96$ (dataset III) respectively, each consisting of $30000$ events. 
In Fig.~6 we show the function $<n_{\pi^+ \pi^-}>_{\epsilon}$ for the CMC datasets I, II and III. The horizontal lines correspond to the 
values $<n_{\pi^+ \pi^-}>(\epsilon)=5$, $<n_{\pi^+ \pi^-}>(\epsilon)=1.5$ and $<n_{\pi^+ \pi^-}>(\epsilon)=1.1$. 
  
\begin{figure}
\includegraphics{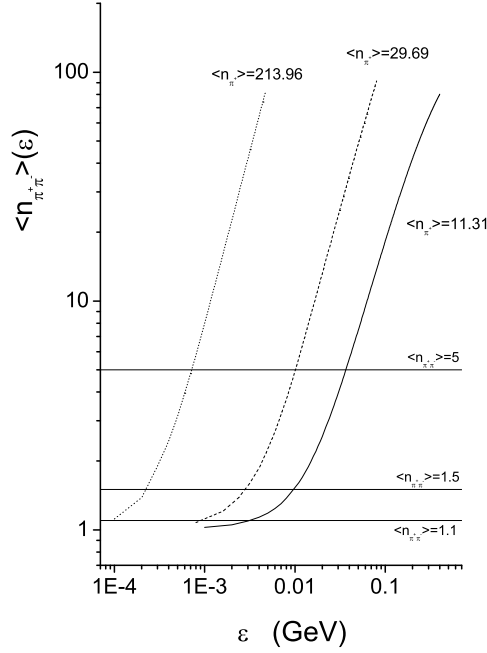}
\caption{\label{fig:fig6} The function $<n_{\pi^+ \pi^-}>(\epsilon)$ for the three different CMC datasets described in text.}
\end{figure}

For increasing size of the system (increasing $<n_{\pi^+}>$) the $\epsilon$ values leading to a given $<n_{\pi^+ \pi^-}>_{\epsilon}$ decrease. 
The complete analysis within our approach for these datasets involves the reconstruction of the isoscalar sector as well as the calculation 
of the corresponding correlator $\Delta F_2$. However a comparative study between the different datasets in terms of factorial moments
is possible only for classes of events characterized by almost the same multiplicity since in the opposite case artificial fluctuations
are induced \cite{WolfDr}. Thus, to compare the systems of different size, we have to calculate the correlator for fixed multiplicity of
reconstructed dipions $<n_{\pi^+ \pi^-}>$ choosing a suitable value of $\epsilon$ in each case. As in the case of dataset I we use the 
effective slope (or critical index) $\phi_2$ of the correlator ($\Delta F_2 \sim M^{2 \phi_2}$) as a measure of the fluctuations 
in the corresponding dataset. 
In Fig.~7 we show the results of our calculations for the three datasets I, II and III using $<n_{\pi^+ \pi^-}>_{\epsilon}=5$ (full circles), 
$<n_{\pi^+ \pi^-}>_{\epsilon}=1.5$ (open stars) and $<n_{\pi^+ \pi^-}>_{\epsilon}=1.1$ (crosses). We plot $\phi_2$ as a function of the 
mean number of positive pions per event $<n_{\pi^+}>$ respectively.       
It is clearly seen in Fig.~7 that the critical fluctuations, within the limitations of our analysis due to statistics, are at best recovered for
CMC systems involving medium to small size nuclei and using in the analysis $\epsilon$ values fullfiling the constraint $<n_{\pi^+ \pi^-}>~<~2$. 
The critical index $\phi_2$ approaches the theoretically expected value $\frac{2}{3}$. In fact we see that for $<n_{\pi^+ \pi^-}>_{\epsilon}=1.1$
and $<n_{\pi^+}>=11.31$ we obtain exactly the critical QCD prediction.
This is the main result of the present work providing us with a useful guide for the search of QCD critical fluctuations 
in relativistic ion collisions.
  
\begin{figure}
\includegraphics{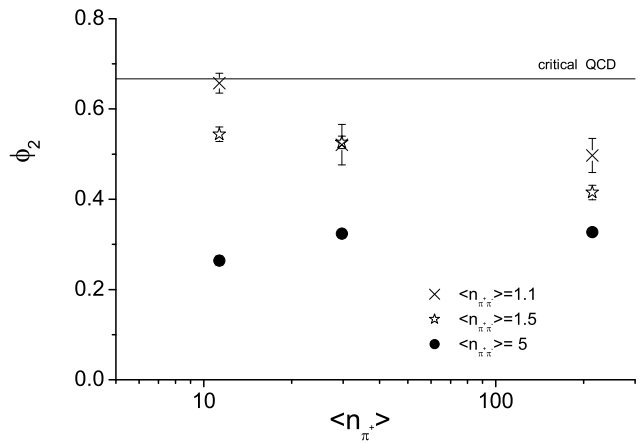}
\caption{\label{fig:fig7} The slope $\phi_2$ as a function of $<n_{\pi^+}>$ for the CMC datasets I, II and III using 
$<n_{\pi^+ \pi^-}>(\epsilon)=5$ (full circles), $<n_{\pi^+ \pi^-}>(\epsilon)=1.5$ (open stars) and $<n_{\pi^+ \pi^-}>(\epsilon)=1.1$ (crosses).}
\end{figure}

\section{Conclusions} 
In conclusion, we have shown that a set of well prescribed observables (factorial moments,
correlators, intermittency exponents) associated with the existence of a critical point in quark matter, 
can be established in nuclear collisions.
These observables belong to the reconstructed isoscalar sector describing massive dipions 
($\pi^+ \pi^-$ pairs) near the two-pion threshold and their behaviour reveals strong critical
effects suggested by $\sigma$-field fluctuations near the critical point. We claim that a 
search for such a critical behaviour in heavy ion experiments is feasible within the 
framework of a reconstruction procedure of the momenta of massive dipions, 
discussed in this work. The critical effects in this sector although independent of the system
size can be at best recovered, through the proposed reconstruction algorithm, in collisions of 
relatively small nuclei. The appropriate kinematical window to look for these effects is also 
determined. Our study has also shown that although it is not possible to reveal any conventional sign
of the sigma itself in experiments with nuclei at high energies, nevertheless its density fluctuations
associated with the critical point are observable and can be measured and studied in a systematic way.
Therefore our proposal is to study, using the above observables, different 
processes at the SPS and RHIC with the aim to scan a substantial area of the phase diagram, 
in a systematic search for the QCD critical point in collisions of nuclei.

\begin{acknowledgments}
We thank the NA49 Collaboration and especially K. Perl and R. Korus for their help in the
treatment of the HIJING data for the $C+C$ system. One of us (N.G.~A.) wishes to thank Larry
McLerran and Dmitri Kharzeev for the kind hospitality at the Brookhaven National Laboratory. This work was 
supported in part by the EPEAEK research funding program PYTHAGORAS (70/3/7315) and also by the research funding program
of the University of Athens "Kapodistrias". 
\end{acknowledgments}

{}

\end{document}